\documentclass[aps,prl,twocolumn,groupedaddress]{revtex4-1}

\usepackage{graphicx}
\usepackage{booktabs}
\usepackage{amssymb,bm,mathrsfs,bbm,amscd}
\usepackage[tbtags]{amsmath}

\newcommand{\rf}[1]{(\ref{#1})}

\def\etal{{\it et al.}}

\def\re{\hbox{Re}\,}
\def\im{\hbox{Im}\,}
\def\al{\alpha}
\def\kn#1#2{k^{\rm N(8)}_{{#1}{#2}}{}}
\def\xb{\overline{x}}
\def\yb{\, \overline{y}}
\def\zb{\, \overline{z}}

\begin{document}

\title{Combined Search for a Lorentz-Violating Force in Short-Range Gravity
\\ 
Varying as the Inverse Sixth Power of Distance 
} 

\author{Cheng-Gang Shao}
\author{Ya-Fen Chen}
\author{Yu-Jie Tan}
\author{Shan-Qing Yang}\email[E-mail: ]{ysq2011@hust.edu.cn}
\author{Jun Luo}\email[E-mail: ]{junluo@sysu.edu.cn}

\affiliation
{MOE Key Laboratory of Fundamental Physical Quantities Measurement,
Hubei Key Laboratory of Gravitation and Quantum Physics,
PGMF and School of Physics, 
Huazhong University of Science and Technology, 
Wuhan 430074, People's Republic of China} 

\author{Michael Edmund Tobar}\email[E-mail: ]{michael.tobar@uwa.edu.au}

\affiliation
{Department of Physics, University of Western Australia, 
Crawley, WA 6009, Australia}

\author{J.C.\ Long}\email[E-mail: ]{jcl@indiana.edu}
\author{E.\ Weisman}
\altaffiliation[Present address: ]{Department of Physics and Astronomy, 
Northwestern University, Evanston, IL 60208, USA}
\author{V.\ Alan Kosteleck\'y}\email[E-mail: ]{kostelec@indiana.edu}
\affiliation
{Physics Department, Indiana University, Bloomington, Indiana 47405, USA}

\date{August 2018}

\begin{abstract}
Precision measurements of the inverse-square law 
via experiments on short-range gravity
provide sensitive tests of Lorentz symmetry.  
A combined analysis of data from experiments 
at the Huazhong University of Science and Technology and Indiana University 
sets simultaneous limits on all 22 coefficients for Lorentz violation 
correcting the Newton force law as the inverse sixth power of distance.
Results are consistent with no effect at the level of $10^{-12}$~m$^{4}$.
\end{abstract}

\maketitle

Lorentz symmetry, 
the idea that physical laws are unchanged under rotations and boosts, 
is built into both General Relativity (GR) and the Standard Model.  
Although GR provides an impressive description 
of a wide variety of gravitational phenomena,
the successful merger of gravitation and quantum physics
may involve a modification of its foundational principles.
This could produce observable deviations from Lorentz symmetry,
emerging from a unified theory such as strings
\cite{ksp}.

Since no compelling evidence for Lorentz violation (LV) currently exists,
model-independent searches for LV in gravity
play an essential role in testing the foundations of GR.
A powerful model-independent approach 
to describing possible low-energy signals of LV 
is effective field theory 
\cite{ak04},
which is widely adopted for experimental analyses 
studying Lorentz symmetry 
\cite{tables,lvgrreview}.
In the pure-gravity limit,
this approach uses a Lagrange density 
containing the usual Einstein-Hilbert term 
and a series of all observer-scalar terms
involving coefficients contracted 
with gravitational-field LV operators 
of increasing mass dimension $d$.

Precision experiments testing the inverse-square law
at short range provide crucial and specific probes of gravitational properties 
\cite{review},
including tests of Lorentz symmetry in gravity
at submillimeter distances
\cite{lk15,hust15,hust-iu16}.  
Applying the techniques of effective field theory in this context
shows that LV operators can lead to  
direction-dependent corrections to the Newton force
that fall as inverse-square, inverse-fourth, inverse-sixth,
and higher powers of distance
\cite{bk06,bkx15,km17}. 
A complete classification of possible effects is known 
\cite{km18},
but no specific predictions exist for their sizes. 
Moreover, 
many of these corrections are experimentally unexplored,
with even comparatively strong ``countershaded'' LV couplings 
remaining untested to date
\cite{kt09}.
Model-independent experimental analyses 
without preconceived sensitivity expectations
are thus vital in investigating this foundational property of GR.

In the present work,
we perform a combined analysis of data from short-range experiments 
at the Huazhong University of Science and Technology (HUST) 
and Indiana University (IU) 
to complete a model-independent search for LV effects
involving operators of mass dimension $d=8$,
which produce a direction-dependent force
inversely proportional to the sixth power of distance.
Our results are consistent with no effects
at the level of $10^{-12}$~m$^{4}$ 
for all 22 independent coefficients for LV
appearing in the Newton limit,
thereby excluding a short-range LV gravitational force
down to a distance scale of less than a millimeter. 

For $d=8$, 
the LV modification to the Newton potential 
between two test masses $m_{1}$ and $m_{2}$ 
is given in spherical polar coordinates by
\cite{km17}
\begin{equation}
V_{LV}(\vec r) = 
-G\sum\limits_{jm}
{\frac{{m_{1}m_{2}}}{{{r^5}}}} {Y_{jm}}(\theta,\phi)k_{jm}^{{\rm{N(8)lab}}}
\label{eq:vglv}
\end{equation}
in the laboratory frame.  
Here, 
the vector
$\vec{r}=\vec{r}_{1}-\vec{r}_{2}
\equiv (r\cos\phi\sin\theta,r\sin\phi\sin\theta,r\cos\theta)$ 
separates $m_{1}$ and $m_{2}$, 
$j=4$ or 6,
and $m$ is an integer in the range $-j\leq m\leq j$. 
The LV effects are controlled 
by the coefficients $k_{jm}^{{\rm{N(8)lab}}}$,
which are complex numbers with dimensions of length to the fourth power.

The explicit form of the coefficients $k_{jm}^{{\rm{N(8)lab}}}$ 
is frame dependent,
so experimental results must be reported in a specified frame.
In cartesian inertial frames in the vicinity of the Earth,
the coefficients can be taken as constant
\cite{sme}.
The canonical frame used in the literature to present results 
is the Sun-centered frame
with right-handed cartesian coordinates $\left(T,X,Y,Z \right)$
chosen such that $T$ is zero at the 2000 vernal equinox, 
the $X$ axis points from the Earth's position at $T=0$ to the Sun,
and the $Z$ axis is parallel to the Earth's rotation axis
\cite{sunframe}.  
Earth-based laboratories are noninertial due to the Earth's rotation,
so the laboratory-frame coefficients $k_{jm}^{{\rm{N(8)lab}}}$ 
acquire dependence on sidereal time \cite{ak98}.
In standard laboratory cartesian coordinates 
with the $x$ axis pointing to the south, the $y$ axis to the east, 
and the $z$ axis to the local zenith,
the laboratory-frame coefficients 
$k_{jm}^{{\rm N(8)lab}}$ 
can be expressed in terms of time-independent coefficients 
$k_{jm}^{{\rm N(8)}}$
in the Sun-centered frame by the relation
\cite{km17} 
\begin{equation}
k_{jm}^{{\rm N(8)lab}} = 
\sum\limits_{m'} 
{{e^{im'{\omega_\oplus}T_\oplus}}} d_{mm'}^{(j)}(-\chi )
k_{jm'}^{{\rm N(8)}},
\label{eq:wigner}
\end{equation}
where the Earth's boost is treated as negligible.
In this expression,
${\omega_\oplus}\simeq 2\pi /(23~{\rm h}~56~\min)$ 
is the Earth's sidereal frequency
and $T_\oplus \equiv T - T_0$ is the local laboratory sidereal time,
which differs from $T$ by a longitude-dependent offset
\cite{offset}:
$T_0 \simeq -3.2$ h for HUST,
and $T_0 \simeq 10.2$ h for IU. 
Also,
$\chi$ is the laboratory colatitude,
and $d_{mm'}^{(j)}$ are the little Wigner matrices
\cite{wigner}.
The primary goal of the experimental analysis is to measure
the coefficients ${k_{jm}^{{\rm N(8)}}}$ in the Sun-centered frame.

The inverse-fifth corrections to the Newton potential 
imply that experiments testing gravity at short range 
have excellent sensitivity to LV effects.  
For $d=8$, 
the index $m'$ in Eq.~\rf{eq:wigner} 
takes integer values in the range $-6\leq m'\leq 6$, 
so the potential includes components 
up to the sixth harmonic of $\omega_{\oplus}$
and can be expressed as a Fourier series in $T$,
\begin{eqnarray}
V_{\rm LV} (\vec r) &=&
-\frac{G m_1 m_2}{r^5}
\Big( 
c_0 + 
\sum\limits_{m=1}^6 
c_m \cos(m\omega_\oplus T_\oplus) 
\nonumber\\
&&
\hskip 90pt
+ s_m \sin(m\omega_\oplus T_\oplus)
\Big).
\label{eq:vgfs}
\end{eqnarray}
The 13 Fourier amplitudes in this expression are functions of
the 22 independent coefficients ${k_{jm}^{{\rm N(8)}}}$
in the Sun-centered frame. 

\begin{table}
\caption{
\label{fourier}
Expressions for the Fourier amplitudes in Eq.\ \rf{eq:vgfs}.}
\renewcommand\arraystretch{1.5}
\setlength{\tabcolsep}{10pt}
\begin{tabular}{cl}
\hline \hline
Quantity & Expression \\ 
\hline
$	c_0	$&$	\al_1 \kn 40 + \al_2\kn60	$	\\
$	c_2	$&$	\al_3 \re \kn 42 + \al_4 \im \kn 42 	$	\\
$		$&$	\hskip 10pt + \al_5 \re \kn 62 + \al_6 \im \kn 62	$	\\
$	s_2	$&$	\al_4 \re \kn 42 - \al_3 \im \kn 42 	$	\\
$		$&$	\hskip 10pt + \al_6 \re \kn 62 - \al_5 \im \kn 62	$	\\
$	c_4	$&$	\al_7 \re \kn 44 + \al_8 \im \kn 44 	$	\\
$		$&$	\hskip 10pt + \al_9 \re \kn 64 + \al_{10} \im \kn 64	$	\\
$	s_4	$&$	\al_8 \re \kn 44 - \al_7 \im \kn 44	$	\\
$		$&$	\hskip 10pt + \al_{10} \re \kn 64 - \al_9 \im \kn 64	$	\\
$	c_6	$&$	\al_{11} \re \kn 66 + \al_{12} \im \kn 66	$	\\
$	s_6	$&$	\al_{12} \re \kn 66 - \al_{11} \im \kn 66	$	\\
$	c_1	$&$	\al_{13} \re \kn 41 + \al_{14} \im \kn 41 	$	\\
$		$&$	\hskip 10pt + \al_{15} \re \kn 61 + \al_{16} \im \kn 61	$	\\
$	s_1	$&$	\al_{14} \re \kn 41 - \al_{13} \im \kn 41	$	\\
$		$&$	\hskip 10pt + \al_{16} \re \kn 61 - \al_{15} \im \kn 61	$	\\
$	c_3	$&$	\al_{17} \re \kn 43 + \al_{18} \im \kn 43 	$	\\
$		$&$	\hskip 10pt + \al_{19} \re \kn 63 + \al_{20} \im \kn 63	$	\\
$	s_3	$&$	\al_{18} \re \kn 43 - \al_{17} \im \kn 43	$	\\
$		$&$	\hskip 10pt + \al_{20} \re \kn 63 - \al_{19} \im \kn 63	$	\\
$	c_5	$&$	\al_{21} \re \kn 65 + \al_{22} \im \kn 65	$	\\
$	s_5	$&$	\al_{22} \re \kn 65 - \al_{21} \im \kn 65	$	\\
\hline
$	\al_1	$&$	\frac {3}{16 \sqrt{\pi}} (3 - 30 \zb^2 + 35 \zb^4)	$	\\
$	\al_2	$&$	-\frac {1}{32} \sqrt{\frac{13} \pi} (5 - 105 \zb^2 + 315 \zb^4 - 231 \zb^6)	$	\\
$	\al_3+ i \al_4	$&$	-\frac {3}{4} \sqrt{\frac{5}{2\pi}} (\xb + i \yb)^2 (1 - 7 \zb^2)	$	\\
$	\al_5+ i \al_6	$&$	\frac {1}{32} \sqrt{\frac{1365}{\pi}} (\xb + i \yb)^2 (1 - 18 \zb^2 + 33 \zb^4)) 	$	\\
$	\al_7+ i \al_8	$&$	\frac {3}{8} \sqrt{\frac{35}{2\pi}} (\xb + i \yb)^4 	$	\\
$	\al_9+ i \al_{10}	$&$	-\frac {3}{16} \sqrt{\frac{91}{2\pi}} (\xb + i \yb)^4 (1 - 11 \zb^2)	$	\\
$	\al_{11}+ i \al_{12}	$&$	\frac {1}{32} \sqrt{\frac{3003}{\pi}} (\xb + i \yb)^6 	$	\\
$	\al_{13}+ i \al_{14}	$&$	-\frac {3}{4} \sqrt{\frac{5}{\pi}} (\xb - i \yb) \zb (3 - 7 \zb^2)	$	\\
$	\al_{15}+ i \al_{16}	$&$	\frac {1}{8} \sqrt{\frac{273}{2\pi}} (\xb - i \yb) \zb (5 - 30 \zb^2 + 33 \zb^4)	$	\\
$	\al_{17}+ i \al_{18}	$&$	\frac {3}{4} \sqrt{\frac{35}{\pi}} (\xb - i \yb)^3 \zb	$	\\
$	\al_{19}+ i \al_{20}	$&$	-\frac {1}{16} \sqrt{\frac{1365}{\pi}} (\xb - i \yb)^3 \zb (3 - 11 \zb^2)	$	\\
$	\al_{21}+ i \al_{22}	$&$	\frac {3}{16} \sqrt{\frac{1001}{\pi}} (\xb - i \yb)^5 \zb	$	\\
\hline \hline
\end{tabular}
\end{table}

Numerical methods can be used to calculate 
the gravitational LV interaction 
between finite test masses.  
Most inverse-square law tests use masses with planar geometry
\cite{hust15-2,yang12}. 
In addition to suppressing the Newton background, 
a planar geometry tends to average 
and suppress the angular oscillations of the LV signal
\cite{hust15,shao16,jcl16},
thereby necessitating careful integration of the forces 
associated with Eq.~\rf{eq:vglv}.  
For practical applications,
it can thus be convenient to calculate 
using a local cartesian coordinate system.  
The spherical harmonics in Eq.~\rf{eq:vglv} can be expanded 
in symmetric trace-free tensors $c_{jm}^{<J>}$ according to
\cite{pw14}
\begin{eqnarray}
{Y_{jm}}(\theta,\phi)=c_{jm}^{*<J> }{n_{<J>}}(x,y,z),
\label{eq:ystf}
\end{eqnarray}
where
\begin{eqnarray}
{n_{<J>}}(x,y,z) = 
\frac{{{r^{j + 1}}}}{{{{( - 1)}^j}(2j - 1)!!}}{\partial _J}\frac{1}{r}.
\label{eq:n_J}
\end{eqnarray}
In this expression, 
${\partial_J}$ represents $\partial_{k_1}\ldots\partial_{k_j}$,
and $c^{<J>}n_{<J>}$ involves a summation 
over all $j$ pairs of repeated indices. 
The tensor $c_{jm}^{<J>}$ is given by
\begin{eqnarray}
c_{jm}^{<J>}=
\frac{{(2j+1)!!}}{{4\pi j!}}\int{{n^{<J>}}} Y_{jm}^*(\theta,\varphi)d\Omega.
\label{eq:stf}
\end{eqnarray}
Applying these results,
the 13 amplitudes in the Fourier series \rf{eq:vgfs} 
can be expressed in terms of cartesian coordinates
and the coefficients ${k_{jm}^{{\rm N(8)}}}$ in the Sun-centered frame.
These expressions are given in Table \ref{fourier}.
The first part of this table displays the 13 amplitudes
in terms of the coefficients ${k_{jm}^{{\rm N(8)}}}$ 
and 22 independent functions $\alpha_{j}(\hat{r},\chi)$, $j=1,\cdots 22$,
of the test mass geometry and the colatitude $\chi$. 
The complex-conjugation relation 
$k_{jm}^{{\rm N(8)}}{}^{*}=(-1)^{m}k_{j(-m)}^{{\rm N(8)}}$
\cite{km09} 
is used to express the ${k_{jm}^{{\rm N(8)}}}$ 
in terms of their real and imaginary parts.  
The functions $\alpha_{j}(\hat{r},\chi)$ 
are specified in the second part of the table,
using the notation
\begin{equation}
\begin{array}{l}
\tilde x = \dfrac xr \cos \chi + \dfrac zr \sin \chi,
\hskip 3pt
\tilde y = \dfrac yr,
\hskip 3pt
\tilde z =  - \dfrac xr \sin \chi + \dfrac zr \cos \chi.
\end{array}
\label{eq:cartro}
\end{equation}
With these results,
it is straightforward to obtain an analytical expression 
for the LV force between a point and finite rectangular plate.
We note that the LV force between a point and an infinite plate vanishes, 
as in the $d=6$ case
\cite{hust15,jcl16}.  
For two finite rectangular plates,
we need merely perform a triple integral to obtain the LV force or torque.

In general, 
measurements of the 13 Fourier amplitudes in a single experiment 
constitute independent signals 
but are insufficient to constrain simultaneously 
the 22 independent coefficients ${k_{jm}^{{\rm N(8)}}}$.  
However, 
two distinct datasets can achieve complete coverage. 
Indeed,
this is true for LV force corrections proportional to $r^{2-d}$, 
for which the number of coefficients is $4d-10$
and the maximum number of signals from any one experiment is $2d-3$.
In the present case with $d=8$,
all 22 coefficients could in principle be measured independently
using two datasets with distinct harmonics from the HUST-2015 experiment 
or using two datasets from the IU-2002 and IU-2012 experiments.
Here,
to maximize the sensitivity to the coefficients ${k_{jm}^{{\rm N(8)}}}$, 
we perform a combined analysis of these four datatsets. 

Details of the HUST-2015 experiment are provided in Ref.~\cite{hust15-2}.
A brief summary is provided here.
A bilaterally symmetric $I$-shaped pendulum is suspended 
near an attractor disk with eightfold symmetry.  
Two planar tungsten test masses of thickness $\sim$200~$\mu$m,
together with two additional tungsten plates slightly offset 
to compensate the Newton torque from $r^{-2}$ interactions,
are mounted on either end of the pendulum facing the attractor.  
The attractor consists of eight similar tungsten source plates 
alternating with eight compensation plates.  
The centers of the attractor and pendulum are aligned 
and the gap between the test and source plates is maintained at 295~$\mu$m.  
The pendulum twist is controlled by a feedback system, 
with differential voltages applied to two capacitive actuators 
on the pendulum.  
In the presence of a non-Newton interaction, 
rotating the attractor produces a torque. 
The attractor rotates at frequency $f_{0}=2\pi /(3846.12s)$, 
so the nominal signal torque oscillates at $8f_{0}$ 
and is well separated from the drive frequency, 
effectively suppressing vibrational backgrounds.  
The experiment is designed to produce approximate null measurements 
by double compensating for both the test and source masses. 

For a Yukawa-type interaction, 
the torque is maximal when the source and test masses are face to face 
and is minimal when they are offset.  
However, 
the LV interaction averages to zero for symmetric configurations
\cite{hust15,jcl16},
so significant contributions appear at the higher harmonics 
$16f_{0}$, $24f_{0}$, $\ldots$.
For the $d=6$ case studied earlier
\cite{hust-iu16}, 
in which the LV signal varies as $r^{-4}$ 
and is well nulled by the compensation scheme,
the $16f_{0}$ signal exceeds the $8f_{0}$ one by an order of magnitude
and only the $16f_{0}$ data were used for the analysis.
In contrast,
the $d=8$ interaction of interest here varies as $r^{-6}$
and is less well nulled, 
so the $8f_{0}$ and $16f_{0}$ contribute about equally.  
The $d=8$ signals at higher harmonics are comparable,
but they are swamped by higher-level noise in the data
\cite{hust15-2},
so we use only the $8f_{0}$ and $16f_{0}$ components 
in the present analysis. 

The LV signal torque in the HUST-2015 experiment 
can be expressed as
\begin{equation}
{\tau _{\rm LV}} =
C_0 + \sum\limits_{m = 1}^6 
{C_m}\cos (m{\omega_\oplus}{T_\oplus}) 
+ {S_m}\sin (m{\omega_\oplus}{T_\oplus}),
\label{eq:taulv}
\end{equation}
where the Fourier amplitudes $C_{m}$, $S_{m}$ 
can be obtained by integration of the amplitudes $c_{m}$, $s_{m}$ 
appearing in Eq.~\rf{eq:vgfs} and Table \ref{fourier}.
This effectively replaces the functions
$\alpha_{j}(\hat{r},\chi)$ with transfer coefficients $\Lambda_{j}$,
defined as
\begin{eqnarray}
{\Lambda_j} = 
G{\rho_1}{\rho _2}\iint 
{\frac{\partial }{{\partial \theta }}
\frac{{{\alpha_j}(\hat r,\chi )}}{{{r^5}}}
d{V_1}d{V_2}},
\label{eq:intalpha}
\end{eqnarray}
in analogy with Eq.~(25) of Ref.~\cite{shao16} for the $d=6$ case.  
For example,
integrating the first row of Table \ref{fourier} via this procedure 
yields $C_{0} = \Lambda_1 \kn40 + \Lambda_2 \kn60$.
The integration \rf{eq:intalpha}
computes the change in torque on the pendulum 
as the source and compensation plates on the attractor 
are swept across the faces of the test and compensation plates 
on the pendulum, 
obtaining the LV torques $\tau_{{\rm LV},8}$ and $\tau_{{\rm LV},16}$ 
at the $8f_{0}$ and $16f_{0}$ response frequencies 
of the pendulum.  
The numerical results for the transfer coefficients $\Lambda_{j}$ 
for both frequencies are listed 
in the second and third columns of Table~\ref{tab:Lambda}.  
The uncertainty on all $\Lambda_{j}$ is $10^{-6}$~Nm/m$^{4}$. 

\begin{table}[!t]
\caption{\label{tab:Lambda} 
Transfer coefficients $\Lambda_{j}$ for HUST-2015, IU-2002, 
and IU-2012 experiments. Errors are 1$\sigma$.}
\begin{ruledtabular}
\begin{tabular}{lrrrr}
Coef-          & HUST 8${f_0}$      & HUST 16${f_0}$ & IU-2012        & IU-2002\\      
ficient&\multicolumn{2}{c}{($\pm 0.01$, $10^{-4}$~Nm/m$^{4}$)} &\multicolumn{2}{c}{($10^{-4}$~N/m$^{4}$)}\\
\hline
$\Lambda_{1}$  & $-0.08$ & $-0.11$ & $92   \pm 269$ & $8   \pm 15$\\ 
$\Lambda_{2}$  & $0.03 $ & $0.14$  & $75   \pm 160$ & $41  \pm 10$\\ 
$\Lambda_{3}$  & $-0.22$ & $0.35$  & $-92  \pm 289$ & $-5  \pm 19$\\ 
$\Lambda_{4}$  & $0.00 $ & $0.00$  & $26   \pm 264$ & $21  \pm 24$\\ 
$\Lambda_{5}$  & $0.22 $ & $0.13$  & $-75  \pm 180$ & $16  \pm 24$\\ 
$\Lambda_{6}$  & $0.00 $ & $0.00$  & $-191 \pm 239$ & $-7  \pm 13$\\ 
$\Lambda_{7}$  & $-0.11$ & $-0.10$ & $-290 \pm 275$ & $4   \pm 25$\\ 
$\Lambda_{8}$  & $0.00 $ & $0.00 $ & $13   \pm 168$ & $2   \pm 19$\\ 
$\Lambda_{9}$  & $0.31 $ & $0.10$  & $642  \pm 512$ & $-48 \pm 28$\\ 
$\Lambda_{10}$ & $0.00 $ & $0.00$  & $-92  \pm 139$ & $-36 \pm 14$\\ 
$\Lambda_{11}$ & $0.09 $ & $-0.02$ & $57   \pm 255$ & $11  \pm 23$\\ 
$\Lambda_{12}$ & $0.00 $ & $0.00$  & $-70  \pm 256$ & $6   \pm 13$\\ 
$\Lambda_{13}$ & $-0.12$ & $0.38$  & $-35  \pm 301$ & $24  \pm 21$\\ 
$\Lambda_{14}$ & $0.00 $ & $0.00$  & $132  \pm 203$ & $15  \pm 12$\\ 
$\Lambda_{15}$ & $0.10 $ & $0.30$  & $178  \pm 319$ & $14  \pm 21$\\ 
$\Lambda_{16}$ & $0.00 $ & $0.00$  & $70   \pm 149$ & $27  \pm 20$\\ 
$\Lambda_{17}$ & $-0.20$ & $0.30$  & $237  \pm 352$ & $6   \pm 14$\\ 
$\Lambda_{18}$ & $0.00 $ & $0.00$  & $-145 \pm 269$ & $3   \pm 22$\\ 
$\Lambda_{19}$ & $0.31 $ & $-0.13$ & $-496 \pm 332$ & $-12 \pm 15$\\ 
$\Lambda_{20}$ & $0.00 $ & $0.00$  & $52   \pm 302$ & $-18 \pm 38$\\ 
$\Lambda_{21}$ & $0.21 $ & $-0.02$ & $-127 \pm 140$ & $-5  \pm 17$\\ 
$\Lambda_{22}$ & $0.00 $ & $0.00$  & $307  \pm 451$ & $52  \pm 11$\\ 
\end{tabular}
\end{ruledtabular}
\end{table}

In the IU-2002 and IU-2012 experiments,
the test masses consist of two planar tungsten oscillators 
of approximate thickness 250~$\mu$m,
separated by a gap of about 80~$\mu$m 
and with a stiff conducting shield between them to suppress backgrounds.
A schematic is given in Fig.~1 of Ref.\ \cite{lk15},
while details of the IU-2002 geometry are given 
in Refs.\ \cite{jcl02,jcl03}
and of the IU-2012 geometry 
in Ref.\ \cite{lk15}.  
The active ``source'' mass drives the force-sensitive ``detector'' mass 
at a resonance near 1~kHz.
At this frequency,
a simple passive isolation system with high bending stiffness
can be used for vibration isolation.  
The oscillations of the detector mass are detected 
using capacitive transducers coupled to a differential amplifier
\cite{yan14}.
The signal is passed to a lock-in amplifier 
referenced by the waveform driving the source mass,
and the output is taken as the raw experimental data 
\cite{lk15}.  
Comparison with the detector thermal noise
permits these data to be converted to force readings.
Details of the IU-2002 calibration are given
in Refs.\ \cite{jcl02,jcl03}
and of the IU-2012 calibration in Refs.\ \cite{yan14,lk15}.  

Following Ref.~\cite{lk15}, 
the theoretical LV force for the IU experiments
is evaluated by Monte Carlo integration 
of the $z$ component of the force from the potential \rf{eq:vglv}, 
incorporating the test-mass curvatures and mode shapes.  
The results can be expressed as a Fourier series 
in the local sidereal time $T_\oplus$ 
analogous to Eq.~\rf{eq:taulv}.
The Fourier force amplitudes are linear combinations 
of the $k_{jm}^{{\rm N(8)}}$,
weighted by a corresponding transfer coefficient $\Lambda_{j}$ 
as in Eq.~\rf{eq:intalpha}.  
The numerical values of the $\Lambda_{j}$ 
for the IU-2002 and IU-2012 experiments
are shown in the fourth and fifth columns of Table~\ref{tab:Lambda}.  
Systematic errors associated with the positions and dimensions 
of the test masses
are established by calculating the mean and standard deviation 
of a population of Fourier amplitudes 
generated with a spread of geometries based on the metrology errors
\cite{jcl02,lk15}.
Many $\Lambda_{j}$ values in all columns of Table~\ref{tab:Lambda} 
are dominated by the error. 
For the IU experiments, 
the error is particularly sensitive to the longitudinal position 
of the detector mass relative to the source mass.       

\begin{figure}
\includegraphics[width=\hsize]{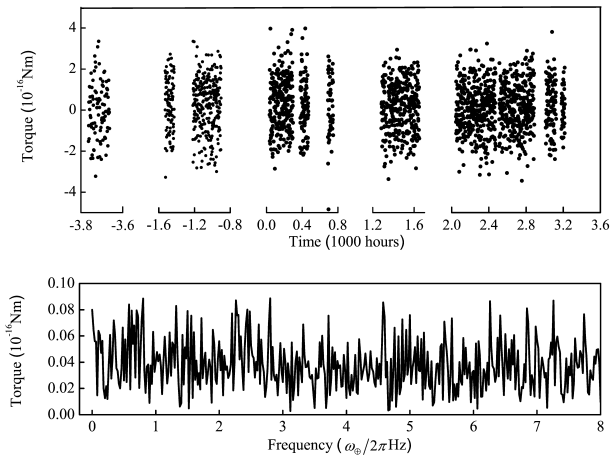}
\vskip-10pt
\caption{\label{fig:hust8} 
HUST-2015 data at $8 f_{0}$ and Fourier transform.} 
\end{figure} 

For the HUST-2015 experiment,
extraction of the LV signal from the data proceeds as described in
Ref.~\cite{hust-iu16}.  
The data rate is much faster than the attractor modulation frequency, 
so data are partitioned into bins corresponding 
to the modulation period $\Delta T=3846.12$~s.  
The LV torque signals $\tau_{{\rm LV},n}(T_\oplus)$ with $n=8$ and 16
are extracted by fitting the measured torque $\tau^{z}(T_\oplus)$ 
in each bin to
\begin{eqnarray}
{\tau^z}(T_\oplus) = \sum\limits_{n = 8,16} {{\tau_{{\rm LV},n}}(T_\oplus)} 
\cos (2\pi n{f_0}T_\oplus + {\varphi_n}),
\label{eq:taumeas}
\end{eqnarray}
where ${\varphi_{n}}$ is set by operation of the experiment.  
The values of $\tau_{{\rm LV},n}(T_\oplus)$ are taken 
to be approximately constant in each bin,
since ${\omega_\oplus}\Delta T \ll 1$ 
and any sidereal variation within each bin is negligible.  
Data for the torque ${\tau_{{\rm LV},8}}$ 
are plotted in the upper panel of Fig.~\ref{fig:hust8} 
as a function of time.
Each point shows the mean measurement in the modulation period
without errors,
which are dominated by statistical fluctuations.  
The Fourier spectrum for these data is displayed 
in the lower panel of Fig.~\ref{fig:hust8}.
The corresponding plots for the torque $\tau_{LV,16}$ 
appear in Fig.~1 of Ref.~\cite{hust-iu16}.

\begin{table}[htb]
\caption{\label{tab:ft} 
Fourier amplitudes ($2\sigma$, 
units $10^{-16}$~Nm for HUST and $10^{-16}$~N for IU).}
\begin{ruledtabular}
\begin{tabular}{lrrrr}
Mode      & HUST-8$f_{0}$   & HUST-16$f_{0}$  & IU-2012      & IU-2002\\        
\hline
${C_0}$  &$0.08  \pm 0.10$ & $-0.20   \pm 2.40 $ & $0   \pm 136$ & $2\pm 411   $\\  
${C_2}$  &$0.00  \pm 0.08$ & $ - 0.01 \pm 0.08 $ & $  47\pm 166$ & $ -53\pm 556$\\
${S_2}$  &$-0.06 \pm 0.08$ & $ - 0.08 \pm 0.08 $ & $-192\pm 187$ & $ -51\pm 176$\\  
${C_4}$  &$0.00  \pm 0.08$ & $ 0.04   \pm 0.08 $ & $ -42\pm 156$ & $ 25\pm 448 $\\  
${S_4}$  &$0.01  \pm 0.08$ & $ - 0.03 \pm 0.08 $ & $ -58\pm 192$ & $ 83\pm 237 $\\ 
${C_6}$  &$0.04  \pm 0.08$ & $ -0.04  \pm 0.08 $ & $ -41\pm 179$ & $ 61\pm 306 $\\  
${S_6}$  &$0.00  \pm 0.08$ & $  0.02  \pm 0.08 $ & $  91\pm 146$ & $ 52\pm 241 $\\  
${C_1}$  &$-0.03 \pm 0.08$ & $ 0.00   \pm 0.08 $ & $-108\pm 193$ & $ 30\pm 130 $\\  
${S_1}$  &$0.03  \pm 0.08$ & $  0.00  \pm 0.08 $ & $   3\pm 161$ & $-192\pm 449$\\  
${C_3}$  &$0.00  \pm 0.08$ & $ 0.01   \pm 0.08 $ & $-173\pm 145$ & $ 215\pm 180$\\ 
${S_3}$  &$0.03  \pm 0.08$ & $ -0.06  \pm 0.08 $ & $ 223\pm 207$ & $ -56\pm 390$\\  
${C_5}$  &$0.02  \pm 0.08$ & $ -0.03  \pm 0.08 $ & $ 142\pm 181$ & $ -98\pm 201$\\  
${S_5}$  &$-0.08 \pm 0.08$ & $  0.05  \pm 0.08 $ & $ 132\pm 165$ & $-190\pm 290$\\  
\end{tabular}
\end{ruledtabular}
\end{table}

\begin{table}[!t]
\caption{\label{tab:pg_III} 
Independent coefficient values
(2$\sigma$, units 10$^{-13}$~m$^{4}$)
obtained by combining HUST and IU data.}
\begin{ruledtabular}
\begin{tabular}{lcrr}
& Coefficient  & Measurement &\\
\hline
& $\kn 40$    & $-6.4 \pm 50.9$ & \\ 
& \re$\kn 41$ & $1.7  \pm 5.5$ & \\ 
& \im$\kn 41$ & $0.9  \pm 5.8$ & \\ 
& \re$\kn 42$ & $0.0  \pm 3.9$ & \\ 
& \im$\kn 42$ & $0.9  \pm 4.0$ & \\ 
& \re$\kn 43$ & $4.3  \pm 7.3$ & \\ 
& \im$\kn 43$ & $2.4  \pm 7.3$ & \\ 
& \re$\kn 44$ & $-2.8 \pm 14.5$ & \\ 
& \im$\kn 44$ & $-2.9 \pm 14.4$ & \\ 
& $\kn 60$    & $5.1  \pm 100.9$ & \\ 
& \re$\kn 61$ & $-2.4 \pm 5.9$ & \\ 
& \im$\kn 61$ & $-1.2 \pm 6.4$ & \\ 
& \re$\kn 62$ & $1.9  \pm 5.5$ & \\ 
& \im$\kn 62$ & $1.7  \pm 6.2$ & \\ 
& \re$\kn 63$ & $4.7  \pm 6.8$ & \\ 
& \im$\kn 63$ & $0.6  \pm 7.9$ & \\ 
& \re$\kn 64$ & $-0.9 \pm 6.8$ & \\ 
& \im$\kn 64$ & $-0.9 \pm 6.7$ & \\ 
& \re$\kn 65$ & $1.2  \pm 7.8$ & \\ 
& \im$\kn 65$ & $3.7  \pm 7.1$ & \\ 
& \re$\kn 66$ & $5.7  \pm 14.4$ & \\ 
& \im$\kn 66$ & $0.9  \pm 14.2$ & \\ 
\end{tabular}
\end{ruledtabular}
\end{table}

The Fourier amplitudes $C_{m}, S_{m}$ are obtained 
by a subsequent fit of the $\tau_{{\rm LV},n}(T)$ data to Eq.~\rf{eq:taulv},
including a small correction for averaging over $\Delta T$
\cite{hust-iu16}.
The results are shown in the second and third columns of Table~\ref{tab:ft}.  
A residual Newton torque is subtracted 
from the time-independent amplitude $C_{0}$.  
The error on this amplitude is dominated by the uncertainties 
on the calculated Newton torque
\cite{hust15-2}, 
which in turn arise primarily 
from uncertainties in the dimensions and positions of the test masses.  
The Newton torque and its error 
are considerably larger for the $16f_{0}$ component, 
which is less well nulled by the compensation scheme.  
The sidereal-harmonic amplitudes in Table~\ref{tab:ft} 
are dominated by the statistical uncertainty, 
which is at the same level for each harmonic.      

For the the IU-2002 and IU-2012 experiments,
the acquired force data are described in detail in Ref.~\cite{lk15}.  
The corresponding Fourier amplitudes up to the sixth harmonic 
of the sidereal frequency $\omega_{\oplus}$ 
are listed in Table~\ref{tab:ft}.  
Uncertainties are dominated by the statistical errors in the data.  
Errors also include contributions from the calibration
\cite{jcl02,lk15} 
and from corrections due to discontinuities in the time-series data
\cite{lk15}, 
the latter of which include here contributions 
from the $5\omega_{\oplus}$ and $6\omega_{\oplus}$ terms 
and hence display slight difference 
relative to the amplitudes reported in Ref.~\cite{hust-iu16}.  
Note that a few modes at $2\omega_{\oplus}$ and $3\omega_{\oplus}$ 
seem to reveal potential resolved signals,
but these subsequently become swamped by geometrical uncertainties 
of the transfer coefficients during the analysis 
and hence yield final measurements of $k_{jm}^{{\rm N(8)}}$ 
consistent with zero.

With the results in Table~\ref{tab:ft} in hand,
the joint analysis proceeds as described
in Refs.~\cite{lk15} and~\cite{hust-iu16}.  
A global probability distribution $P(\bm{\tilde{f}}|\bm{k})$ 
is formed using the 52 Fourier amplitudes 
$\tilde{f}_{i}$ in Table~\ref{tab:ft} and their errors. 
Each measured amplitude is assigned a gaussian distribution $p_{i}$ 
that is a function of the 22 independent $k_{jm}^{{\rm N(8)}}$
and has mean $\mu_{i}$ and standard deviation $\sigma_{i}$.  
The product of the individual $p_{i}$ defines the global distribution,
\begin{equation}
P(\bm{\tilde{f}}|\bm{k}) = 
P_{0}\exp\left[-\sum_{i=1}^{52}
\frac{(\tilde{f_{i}}-\mu_{i})^{2}}{2\sigma^{2}_{i}}\right],
\end{equation}
where $P_{0}$ is an arbitrary normalization.  
The predicted signal $\mu_{i}$ for the $i$th amplitude 
is given by the appropriate Fourier component 
for the HUST or IU experiments,
with the function $\alpha_{j}$ 
replaced by the associated integrated transfer coefficient $\Lambda_{j}$
in Table~\ref{tab:Lambda}.  
The variance $\sigma_{i}^{2}$ incorporates 
all statistical and calibration errors.  
Following standard procedure
\cite{pdg}
to account for the metrology errors on the $\mu_{i}$, 
the global distribution is replaced with the expression
\begin{equation}
P^{\prime} (\bm{\tilde{f}}|\bm{k}) =
\int P(\bm{\tilde{f}}|\bm{k},\bm{x})\pi(\bm{x})d\bm{x},
\label{eq:Pp}
\end{equation}
where $\bm{x}$ represents the set of geometry variables 
and $\pi(\bm{x})$ is their prior probability density function.  
For simplicity, 
for each geometry parameter $x$, 
$\pi(x)$ is taken to be a uniform distribution centered at the measured $x$ 
with a width of twice the error $\Delta x$, 
so that the integral \rf{eq:Pp} reduces to an average over $\bm{x}$.
Independent measurements of each component $k_{jm}^{{\rm N(8)}}$ 
are then obtained by integrating $P^{\prime} (\bm{\tilde{f}}|\bm{k})$ 
over all other components.  
The result is a distribution for the chosen component 
with a single mean and standard deviation, 
which constitute the estimated component measurement and its error.  

Table~\ref{tab:pg_III} displays 
the final results obtained from this joint HUST-IU analysis
for the 22 independent coefficients $k_{jm}^{{\rm N(8)}}$ for LV 
in the Sun-centered frame.
The results are consistent with no LV force 
varying according to the inverse sixth power,
at the level of ${10^{-12}}$~m$^{4}$.  
These measurements are the first of their kind,
and they set a benchmark excluding short-range LV gravitational forces
down to a distance scale of below a millimeter.
They thereby enhance the scope of recent constraints
on LV operators in pure gravity with 
$d=4$ 
\cite{bk06,2007Battat,2007MullerInterf,2009Chung,jcl10,2010Panjwani,%
2010Altschul,2011Hohensee,2012Iorio,2013Bailey,2014Shao,%
kt15,he15,bo16,yu16,pl16,bo17,2016Flowers,2017Abbott,2017Shao}
$d=5$
\cite{km16,2018Bailey},
$d=6$
\cite{kt15,lk15,hust15,hust-iu16,km17,yu16,km16}
$d=7$
\cite{km16},
$d=8$
\cite{kt15},
and $d=10$
\cite{jt16}.

This work was supported in part
by the National Natural Science Foundation of China 
under grants No.~91636221, No.~11722542, No.~91736312, and No.~11805074,
by the United States National Science Foundation under grant No.~PHY-1707986, 
by the United States Department of Energy under grant No.~{DE}-SC0010120,
and by the Indiana University Center for Spacetime Symmetries.

\end{document}